%% file: main.tex
\documentclass[review]{elsarticle}

\usepackage{lineno,hyperref}
\journal{Information Processing Letters}

\usepackage{xspace}
\usepackage{amsmath, amssymb,algorithm, algorithmicx,latexsym,amsfonts,amstext} 
\usepackage[noend]{algpseudocode}
\usepackage{listings}
\usepackage{multirow} 
\usepackage{mathtools} % for using the symbol ::

\usepackage{graphicx}
\usepackage{float}
\usepackage[caption=false,font=footnotesize]{subfig}

\algnewcommand{\LineComment}[1]{\State \(\triangleright\) #1}  %inserts comments on 

\def \X {\ensuremath{\mathcal{X}}\xspace}
\def \V {\ensuremath{\mathcal{V}}\xspace}

\def \P {\ensuremath{\mathcal{P}}\xspace}

\DeclarePairedDelimiter\abs{\lvert}{\rvert}% for abs

\newcommand{\degree}{\ensuremath{^\circ}} %command to insert degree symbol
\newcommand{\Reals}[0]{\ensuremath{\mathbb{R}}}

\newtheorem{definition}{Definition}
\newtheorem{theorem}{Theorem}
\newtheorem{lemma}[theorem]{Lemma}

\begin{document}

\begin{frontmatter}

\title{An Efficient Algorithm for Vertex Enumeration of Two-Dimensional Projection of Polytopes} % using Support Functions}

%% Group authors per affiliation:
\author{Amit Gurung* and Rajarshi Ray}
\address{Department of Computer Science \& Engineering,		National Institute of Technology Meghalaya, Shillong - 793003, India \\
Email: amitgurung@nitm.ac.in, rajarshi.ray@nitm.ac.in}

\cortext[mycorrespondingauthor]{Corresponding author}
\ead{amitgurung@nitm.ac.in}

\begin{abstract}
An efficient algorithm to enumerate the vertices of a two-dimensional (2D) projection of a polytope, is presented in this paper. The proposed algorithm uses the support function of the polytope to be projected and enumerated for vertices. The complexity of our algorithm is linear in the number of vertices of the projected polytope and we show empirically that the performance is significantly better in comparison to some known efficient algorithms of projection and enumeration.
\end{abstract}


\end{frontmatter}

%\linenumbers
% ********************************************
		\input{introduce.tex}

		\input{supFunc-Graph.tex}
		
		\input{enumerateMultiModal.tex}
		\input{experiments.tex}
		\input{conclusion.tex}
% ********************************************
\section*{Acknowledgments}
This work was supported by the National Institute of Technology Meghalaya, India and by the DST-SERB, GoI under project grant No. YSS/2014/000623.
\section*{References}
\bibliography{mybiblio}

\end{document}

%% file: introduce.tex
\section{Introduction}\label{sec:intro}

A convex polytope has two common representations, the V-representation and the H-representation\cite{ziegler1995lectures}. In the V-representation, a polytope is represented by the set of its corner points or vertices. In the H-representation, a polytope is represented as an intersection of finitely many half-spaces given as linear inequalities (i.e. Ax $\leq$ b). The vertex enumeration problem is to determine the set of all vertices of a polytope given in the H-representation. In this paper, we address the problem of enumerating the vertices of a two-dimensional projection of a H-represented polytope. A motivation to address this problem is that applications using high dimensional polytopes (in H-representation) may call for its visualization. In such a case, the vertices of the projection of the polytope on two or three dimensions are to be enumerated. This problem may be interesting for applications beyond just visualization, apart from its theoretical relevance. 

The stated vertex enumeration problem has two straightforward solutions. The first is to project the H-represented polytope (to be denoted as H-polytope) using  projection algorithms (Fourier-Motzkin elimination\cite{schrijver1998theory}, Equality Set Projection (ESP)\cite{jones2004equality} etc.) followed by enumerating the vertices of the projection using vertex enumeration algorithms ( LRS\cite{avislrs}, Primal-Dual method\cite{bremner1998primal} etc.). %The complexity of such a solution on a first look appears to be exponential on the dimension of the polytope.  The best algorithm known to us for the projection of a H-polytope, in terms of time complexity, is the ESP algorithm proposed in \cite{jones2004equality}. The complexity of the algorithm is $O(n_f)$, where $n_f$ is the number of inequalities in the projected polytope. The algorithm involves solving a number of linear programs that is linearly dependent on $n_f$. After the projection, vertex enumeration algorithms like the lrs takes time of the order $O(md^2)$ per vertex, where $d$ is the dimension and $m$ is the number of inequalities in the polytope. The number of vertices for a polytope may range from one to $m^{\lfloor d/2 \rfloor}$\cite{avislrs}. %The paper \cite{avis2000estimating} presents a method for estimating the number of vertices of a polytope. 
The time complexity of this approach is the sum of the complexity of projection (best known to us is $O(n_f)$, where $n_f$ is the number of constraints in the projected polytope \cite{jones2004equality}) and enumeration (exponential in the dimension of the projected polytope). The second solution is to first enumerate the polytope vertices and then project on the desired dimensions. The complexity of this approach is the sum of the complexity of vertex enumeration (exponential in the dimension of the polytope) and the complexity of the projection operation that sets all the dimensions other than the projection dimensions to 0, for every enumerated vertex (linear in the number of vertices of the polytope). Overall, the first solution is superior since it is exponential on the dimension of the projected polytope unlike the second, which is exponential on the dimension of the polytope. In this paper, we propose an algorithm which does not explicitly compute the projection (like in the first solution) nor does it enumerate all vertices (like in the second solution). The complexity of our proposed algorithm is linear in the number of vertices of the projected two-dimensional polytope. Since the number of vertices can be exponential in the dimension of the polytope, the complexity of our algorithm is similar to the complexity of the first solution. %We observed that for a $10$-dimensional permutahedron that has $3628800$ vertices, when projected over 2D, has only $6$ vertices which took an insignificant amount of time to enumerate as compared to enumerating all vertices using the lrs\cite{avislrs, avis1998computational} algorithm. 
However, we show empirically that the performance of the proposed algorithm is significantly better in comparison to the first solution of projection followed by enumeration using the algorithms best known to us. In addition, our algorithm generates vertices in a sorted counter-clockwise order. It works on general polytopes unlike vertex enumeration algorithms like \cite{bremner1998primal} that requires the origin at the interior of the polytope. The implementation of the algorithm, the benchmark polytopes for the experiments and instructions for repeatability evaluation can be found in \url{https://bitbucket.org/rajgurung777/vertexenumerationbyprojectionprojects}.

%An immediate application of 2D projection is also seen in the area of reachability analysis to visualize the computed reachable states and also to compare precision of one algorithm with others\cite{DBLP:conf/hybrid/RayG15, DBLP:conf/hvc/RayGDBBG15}.

The paper is organized as follows. In Section \ref{sec:prelims}, we discuss the preliminaries that forms the base of the proposed algorithm. Section \ref{vertEnumSF} discusses the algorithm in detail. In Section \ref{sec:experiments}, we present an empirical study of the performance with the existing competitive algorithms. We conclude in Section \ref{conclude}.

%% file: supFunc-Graph.tex
\section{Preliminaries}\label{sec:prelims}
Our algorithm is based on the use of support functions as a means of representing compact convex sets.

%\subsection{Support Functions}\label{sec:supp-func}
%In this section, we lay down some of the fundamental definitions on support function representation\cite{schneider2013convex, ghosh1998support} for convex set which is used as the set representation for convex n-polytope in our algorithm.
\begin{definition} \cite{ghosh1998support}  Given a nonempty compact convex set $\X \subset \Reals^n$ the \emph{support function} of $\X$ is a function $sup_{\X}:\Reals^n \to \Reals$ defined as:
 \begin{equation}\label{def:sf}
  sup_{\X}(v) = max\{v \cdot x \mid x \in \X\}
 \end{equation}
\end{definition} 
We use the symbol $\rho_i$ to denote the $sup_{\X}(v_i)$ for a vector $v_i$ and a vector $x \in \X$ such that  $v_i \cdot x = \rho_i$ is called the \emph{support vector}, denoted as $\mathcal{SV}(\rho_i)$. The support function  of a compact convex set $\X$ uniquely represents the set $\X$ by the following relation.
 
\begin{equation}\label{equ:convexSet}
	\X = \{x \in \Reals^n | v \cdot x \le sup_{\X}(v) \ for \, all \, v \in S^{n-1} \}
\end{equation}
where $S^{n-1}$ is a unit sphere in $\Reals^n$ and thus $v \in S^{n-1}$ is a unit vector.

\begin{figure}
%\centering{}

	\subfloat[8 facet Polygon\label{fig:poly}]
	{\includegraphics[scale=0.22]{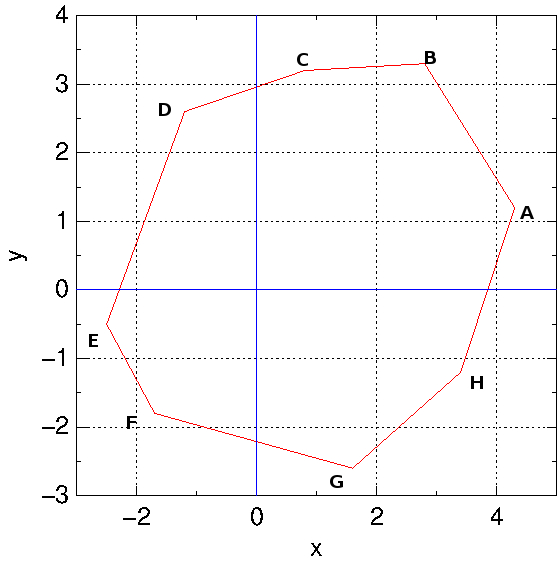}} 
 \hfill
  	\subfloat[Square\label{fig:square}]
  	{\includegraphics[scale=0.22]{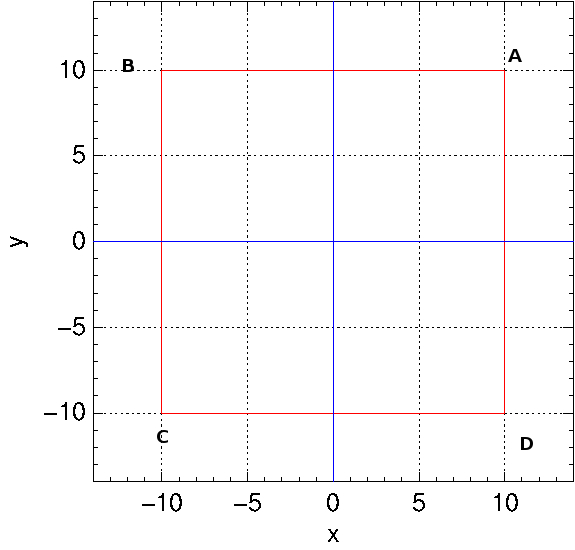}}  
 \hfill
  	\subfloat[Diamond\label{fig:dimond}]
  	{\includegraphics[scale=0.22]{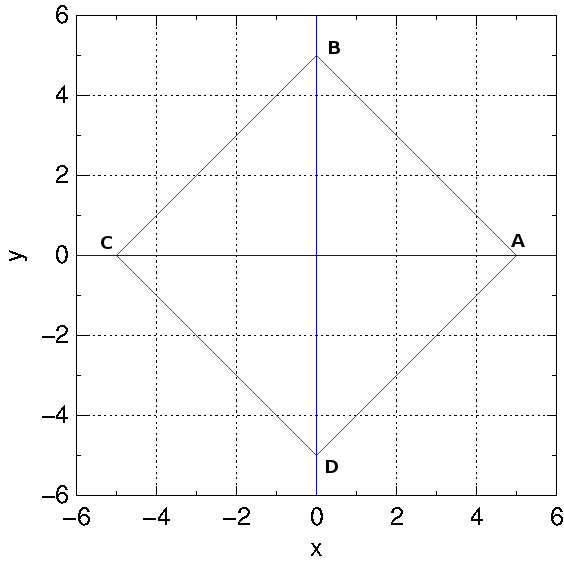}}
  \hfill
	\subfloat[Polygon: ($\Theta_i,\rho_i)$ \label{fig:poly2}]
	{\includegraphics[scale=0.22]{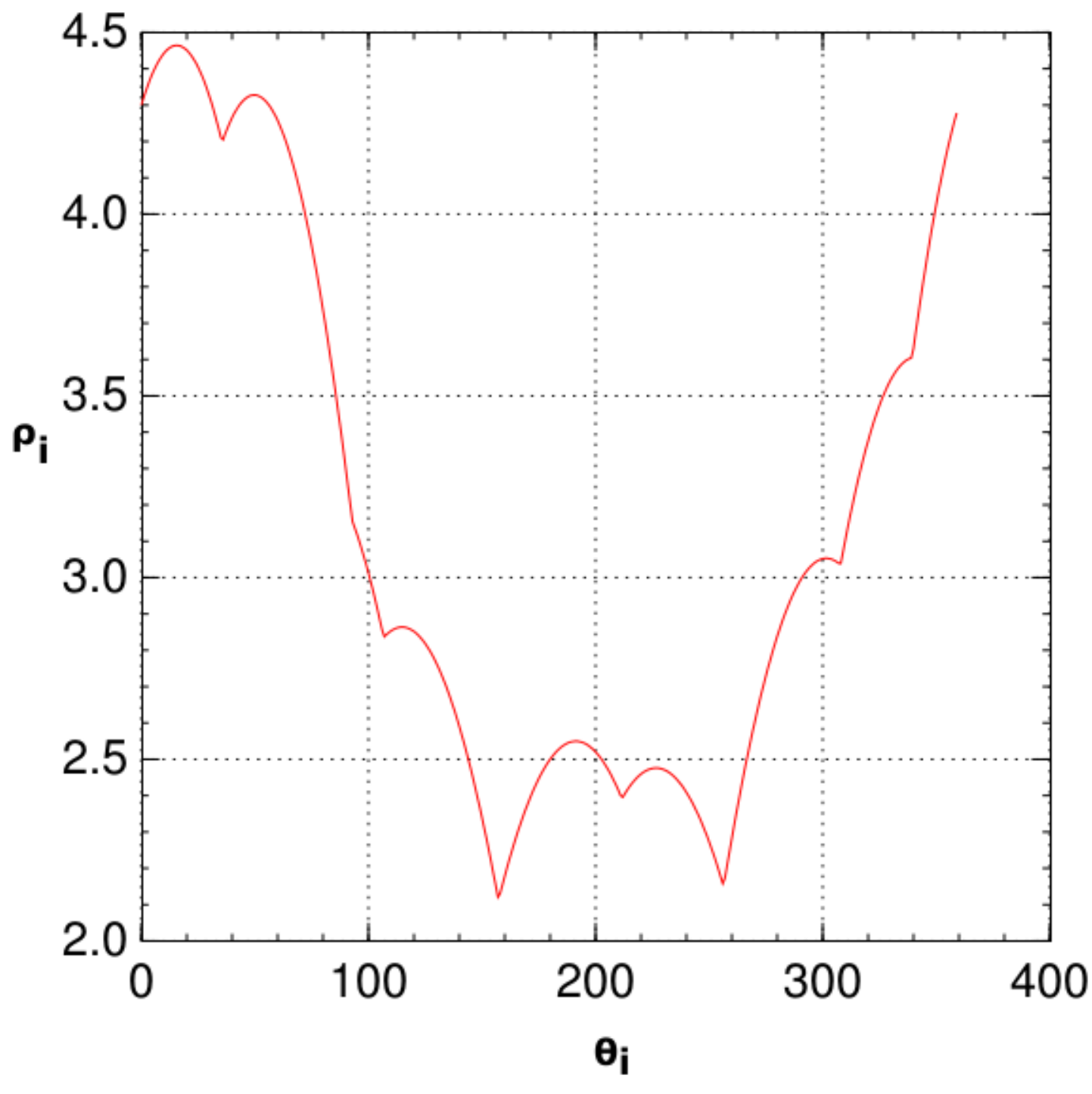}}
 \hfill
 	\subfloat[Square: $(\Theta_i,\rho_i)$ \label{fig:square2}]
 	{\includegraphics[scale=0.22]{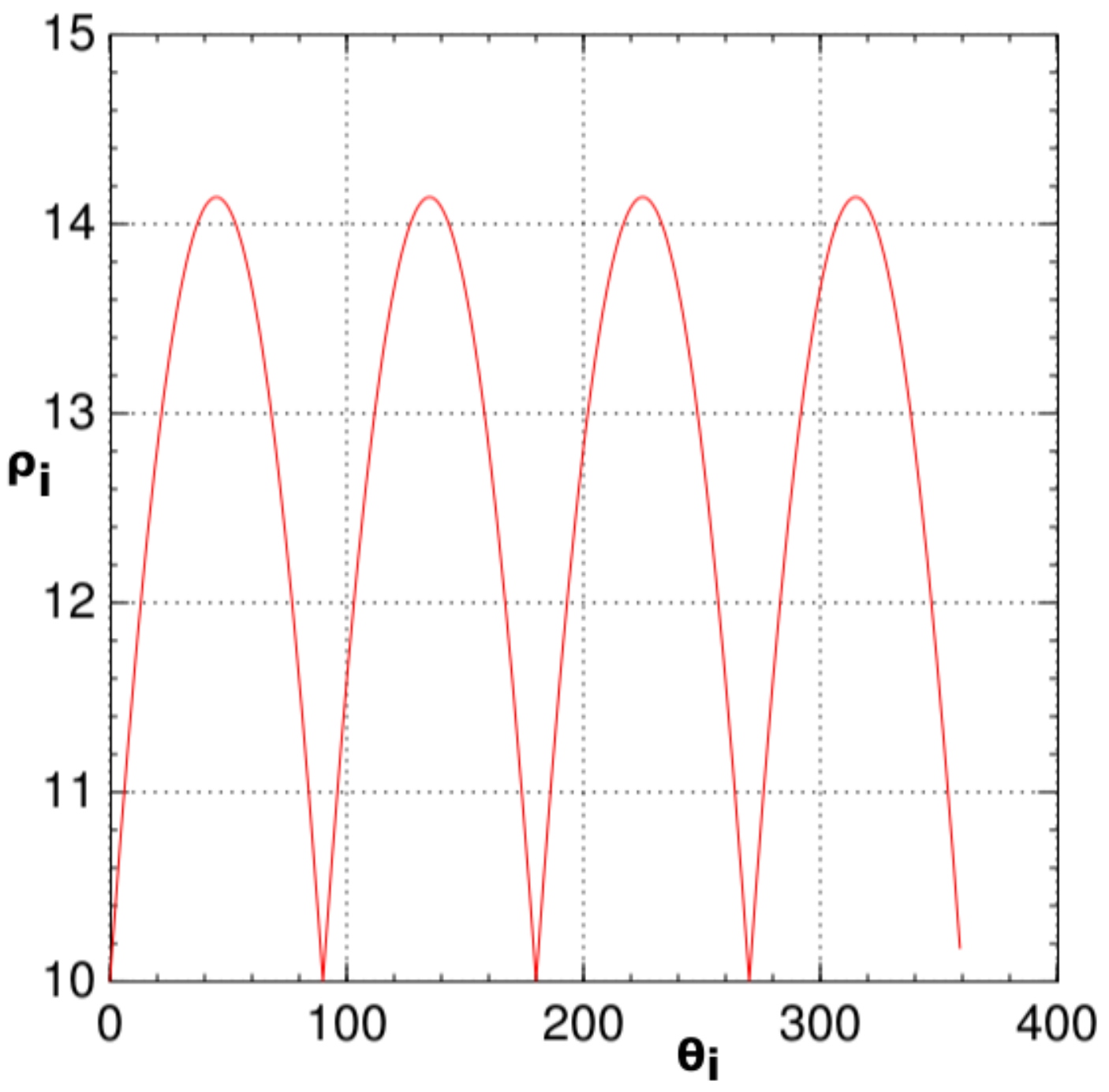}}
 \hfill
 	\subfloat[Diamond: $(\Theta_i,\rho_i)$ \label{fig:dimond2}]
	{\includegraphics[scale=0.22]{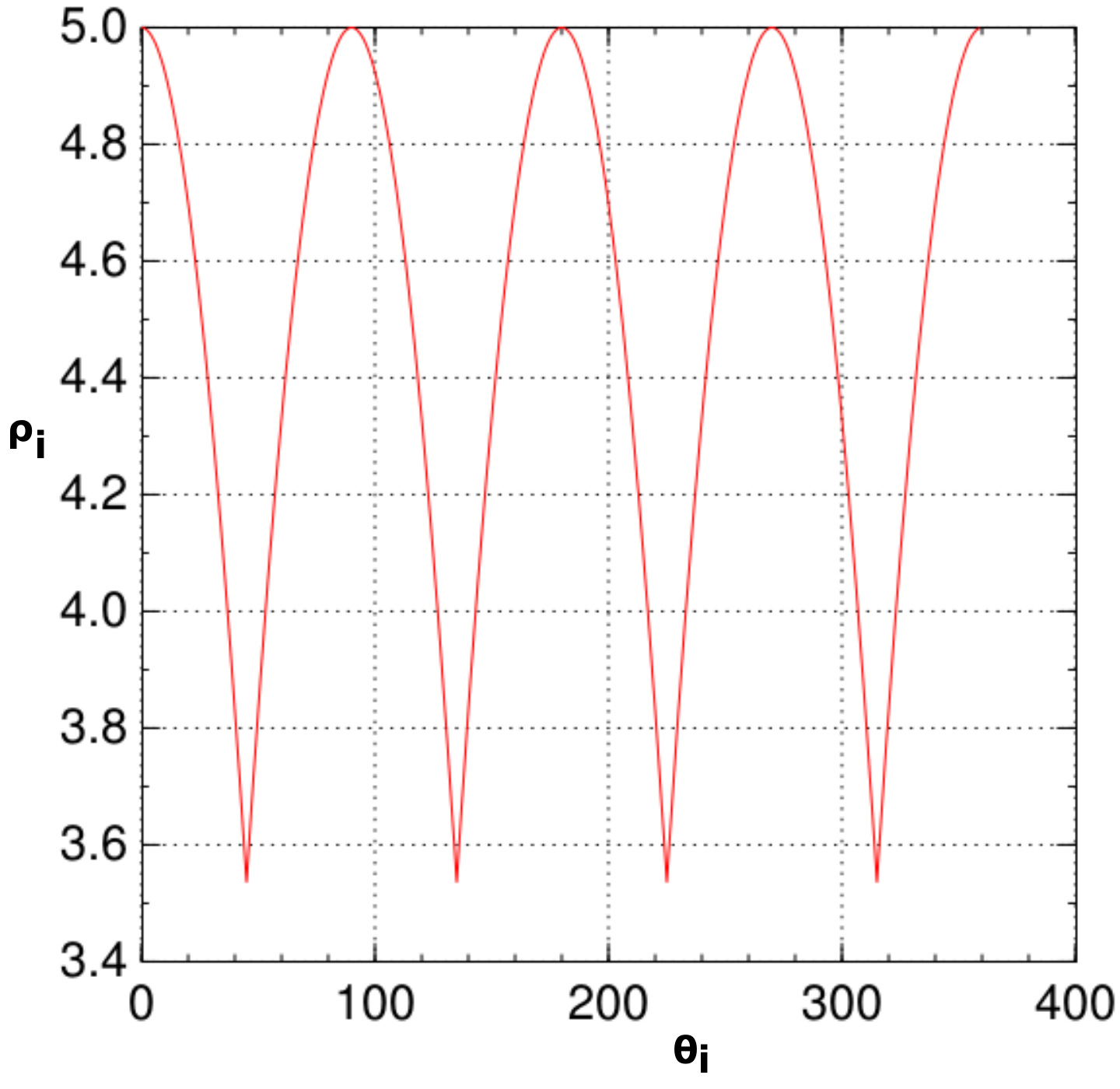}}

	\caption{$\Theta_i$ vs. $\rho_i$ plots of polytopes (a), (b) and (c) are shown in (d), (e) and (f) respectively. \label{fig:theta-rho-space}}
\end{figure}

The pairs $(v_i,\rho_i)$ denote the support function samples of any convex set. In $\Reals^2$, there exists a bijection from the set of angles, say in degrees, $\Theta=\{0, 360\}$ to the set of unit vectors in $S^{1}$. The angle $\theta \in \Theta$ gives a unique orientation of the unit vector, and the bijection is $v=(cos \theta, sin \theta)$. Therefore, the support function of convex sets in $\Reals^2$ can also be represented in the domain of $\Theta$. Note that similar transformation of domain is also defined for convex sets in higher dimensions. The plot of $\Theta_i$ versus $\rho_i$ of a polytope in $\Reals^2$, in particular, gives a continuous collection of sinusoids, as shown in Figure \ref{fig:theta-rho-space}. We now present a lemma giving a relation between the vertices and sides of a 2D-polytope and its support function samples $(\Theta_i,\rho_i)$. 

\begin{lemma}\cite{ghosh1998support}
The support vector $\mathcal{SV}(\rho_i)$ for any $\Theta_i$ corresponding to a sinusoid in the plot of support function samples, is a unique vertex of the polytope. The support vector $\mathcal{SV}(\rho_i)$ for the $\Theta_i$ at a junction of two sinusoids, is a vector of some side of the polytope.
\label{lem:sine-curve}
\end{lemma}
%The application of Lemma \ref{lem:sine-curve} can be found in a similar work in \cite{frehse2012flowpipe} where the intersection of two convex sets reduces to a univariate minimization problem when convex sets are represented as support functions. However, in the context of vertex enumeration, the problem reduces to a multimodal problem requiring to determine more than one solutions, we describe this in detail in the following section.

%% file: enumerateMultiModal.tex
%\newcommand{\degree}{\ensuremath{^\circ}}
\section{Vertex Enumeration using Support Functions}\label{vertEnumSF}
%Our algorithm focus in enumerating the vertices of two-dimensional projection of a H-polytope where the projecting dimensions are supplied by the user. %If we consider $x_1$ and $x_2$ as the projecting variables, then the enumeration problem boils down to the domain of $x_1x_2$ plane. 
%We use the support functions as a set representation for representing the H-polytope as described in Equation \ref{equ:convexSet}. %A vertex enumeration problem of a convex polygon in $(\Theta_i,\rho_i)$ representation, is equivalent to the problem of identifying the number of sinusoidal curves in the given projection $x_1x_2$ plane (Lemma \ref{lem:sine-curve}). Now, the challenging task is to determine the number of these curves (or vertices) and their width i.e., the range of $\Theta$'s that forms each curve as shown in Fig. \ref{fig:vertice-edges}. This problem can be mapped to a multimodal problem.

Our algorithm is based on two key observations. First observation is that the support vectors $\mathcal{SV}(\rho_i)$ corresponding to the sinusoids gives the vertices of the polytope, using lemma 1. Moreover, every vertex of the polytope corresponds to some sinusoid. Therefore, the vertex enumeration problem of 2D-polytopes reduces to finding the support vectors for every sinusoid. The second observation is that for a polytope $\P$ in $\Reals^n$, we can obtain the support function samples of the projection of $\P$ in any two desired dimensions, say ($i$,$j$), by sampling its support functions in $v \in S^{n-1}$ such that $v_k = 0$ for all $k\neq i,j$. In this way, we can obtain the support function samples ($\Theta_i$,$\rho_i$) of a 2D projection of a polytope $\P$ without explicit projection. 

%The $(\Theta_i,\rho_i)$ representation of a convex polygon maps a vertex (from a single point) to a curve (a series/range of points). Thus, to determine a vertex, it is enough to determine a single point from the range of points. %(denoted by the curve as defined above).
%\begin{figure}[!thb]
%\centering{}	
%	\includegraphics[scale=0.22]{fig/Vertices-Edges.jpg}
%	\caption{Vertices as curves and edges as points in $(\Theta_i,\rho_i)$ representation of a Square. $\Theta$ for different vertices ranges from [0-90], [90-180], [180-270] and [270-360] degrees respectively.\label{fig:vertice-edges}}
% \end{figure}

 \begin{figure}
 \centering{}
 
 	\subfloat[$(\Theta_i,\rho_i)$ representation\label{fig:vertice-edges}\label{fig:vertices-range}]
 	{\includegraphics[scale=0.18]{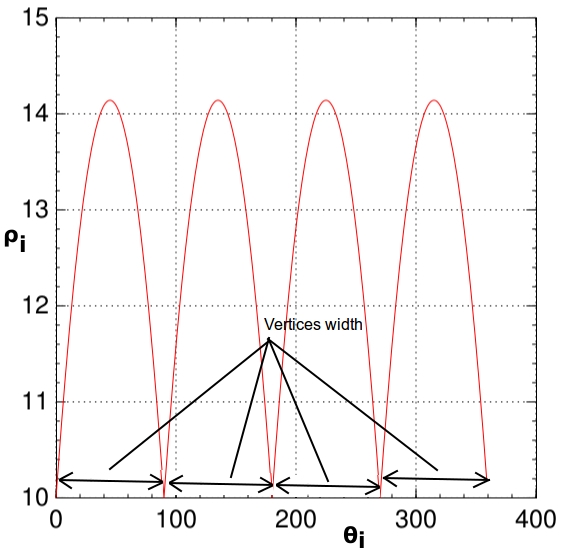}} 
 % \hfill
   	\subfloat[ESP plot in MatLab\label{fig:esp-plot}]
   	{\includegraphics[scale=0.27]{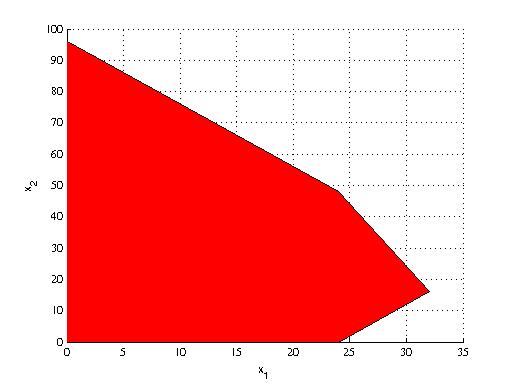}}  
 % \hfill
   	\subfloat[SFA plot with plotutils\label{fig:saf-mit-plot}]
   	{\includegraphics[scale=0.27]{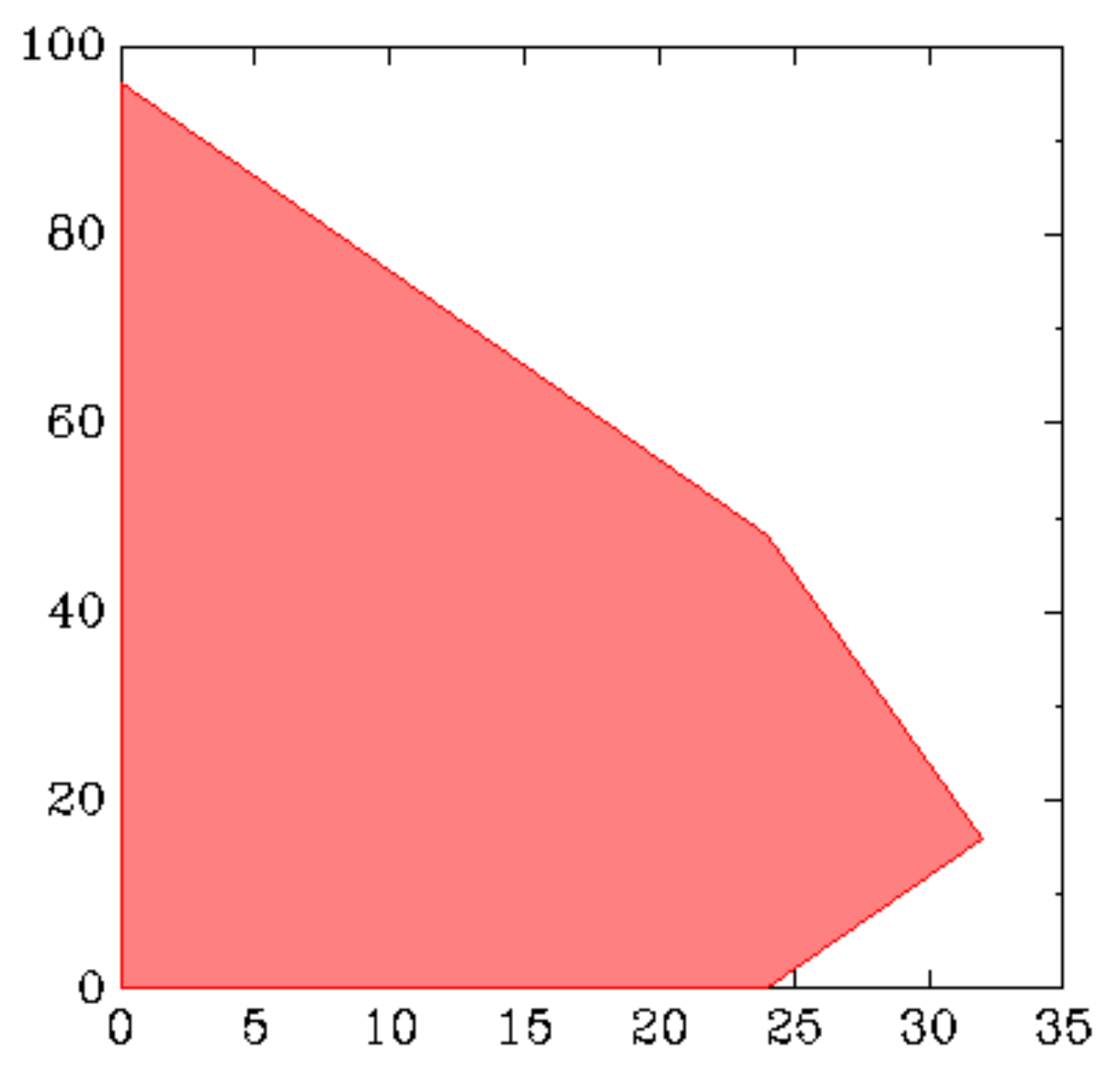}} 
 
 	\caption{(a) Support function samples $(\Theta_i,\rho_i)$ of a Square and indicating the $\Theta_i$ ranges for different vertex as (0-90), (90-180), (180-270) and (270-360) degrees respectively, (b) Output of a projected polytope using the ESP algorithm on the mit benchmark and (c) Output of our vertex enumeration algorithm on the mit benchmark. \label{fig:three-in-one}}
 \end{figure}
\subsection{Vertex Enumeration Algorithm}
 
We denote the $\Theta_i$ at a junction of two sinusoids as a \emph{critical point}. In Fig. \ref{fig:vertice-edges}, the points $\Theta =90\degree$, $180\degree$ and $270\degree$ are the critical points. Algorithm \ref{algo:enumerate} identifies all the critical points, thereby enumerating the vertices of the projection using a variant of binary search procedure. In particular, the algorithm begins by computing the support function $\rho_i$ for $\Theta_i=0\degree$ and obtain its corresponding vertex $p1$ in line \ref{algo:EvalVertex}. Computing the support function of a convex polytope is equivalent to solving a linear program (LP). The computed vertex is added to the data-structure $\V$. By employing a variant of binary search procedure in line \ref{algo:CallBS}, the algorithm iterates to find all critical points and its corresponding immediate vertex (i.e., $NextVertex$) in the domain $\Theta=[0,360]\degree$. The binary search procedure returns the immediate next critical point and its immediate vertex with respect to the start point (the argument $LB$). The critical point becomes the new start point ($LB$) in the next iteration. The new vertex $NextVertex$ is added to the data-structure $\V$. When the start point ($LB$) is equal to the returned critical point, then the corresponding vertex of $LB$ is equal to the returned $NextVertex$, the algorithm terminate. The order of insertion of these vertices is sorted in a counter-clockwise order, because the process of determining critical points progresses from $\Theta=0\degree$ to $360\degree$.
\begin{algorithm}[!thb]
	\caption{Vertex Enumeration of a $2D$ projection of a H-polytope.}
	\textbf{Input arguments:} H-polytope $\P$ in $A.x \leq b$, projecting dimensions $d_1$ and $d_2$ \\
	\textbf{Output:} Enumerated vertices of the projection in $\V$.
	\begin{algorithmic}[1] % with line numbering
		\Procedure{VERTEX-ENUMERATE}{$\P$, $d_1$, $d_2$}
			\State $\Theta_i = LB = 0\degree$; $UB=360\degree$; $done=false$ \label{algo:jump}
			\State $\rho_i \gets sup_{\P}(\Theta_i)$ ; $p_1 \gets \mathcal{SV}(\rho_i)$; Insert $p_1$ in $\V$ \label{algo:EvalVertex} \Comment{Get the first vertex in $\V$}
			\Repeat \label{while-loop}
					\State $ LB = BinSearch(LB, UB, p_1, NextVertex)$ \label{algo:CallBS} 
					\If {$p_1 = NextVertex$} {$done = true$}
					%	\State {$done = true$}
					\EndIf
					\State $ p_1 \gets NextVertex$, insert $NextVertex$ in $\V$ \label{algo:insert2}
	   	 	\Until {$done$}
		\EndProcedure
	\end{algorithmic}\label{algo:enumerate}
\end{algorithm}
\begin{algorithm}[!thb]
	\caption{Procedure to aid the proposed vertex enumeration algorithm}
	\textbf{Input argument:}  We set a desired value to the error tolerance parameter $\epsilon$\\
	\textbf{Output:} Returns the critical point and the immediate vertex $NextVertex$.
	\begin{algorithmic}[1] % with line numbering
		\Procedure{BinSearch}{$LB, UB, SearchVertex, NextVertex$}
			\State $previousMid \gets 0.0$
			\While {$LB \le UB$}\label{while-loop}
		    	\State $mid \gets (LB + UB) \div 2$
		    	
		    	\State $\rho_{mid} \gets sup_{\P}(mid)$ ; $px \gets \mathcal{SV}(\rho_{mid})$;
				\If {($px \ne SearchVertex$)} \Comment{$mid$ is on right side of critical point}
					\State $UB \gets mid$, $NextVertex \gets px$
				\ElsIf {$px = SearchVertex$}\Comment{$mid$ is on left side of critical}
					\State  $LB \gets mid$
				\EndIf				
				\If {$\abs{(mid-previousMid)} < \epsilon$}
					\State \textbf{return $mid$}
				\EndIf
				\State $previousMid \gets mid$
	      	\EndWhile
		\EndProcedure
	\end{algorithmic}\label{algo:binSearch}
\end{algorithm}
\subsection {Procedure Binary Search}
Given the bounds $LB$, $UB$ (lower, upper) and a vertex $SearchVertex$, the goal of the binary search procedure in Algorithm \ref{algo:binSearch} is to return an immediate critical point that lie towards the right of $LB$. Each time the search domain [$LB,UB$] is reduced by half until it converges to the critical point. To avoid numerical inaccuracy, we search a critical point until the search interval size goes less than a tolerance $\epsilon$, which can be assigned as an input to the algorithm. The algorithm assumes that there are no critical points in an interval of size less than or equal to $\epsilon\degree$. The procedure also returns the $NextVertex$ that denotes the immediate vertex corresponding to the critical point. The time complexity of binary search is $O(\log n)$ where $n$ is the distance between the two critical points (or vertices) denoted by a range $[\Theta_1, \Theta_2]\in \Reals$. If we consider $\V$, the number of vertices in the projecting plane, then the first vertex is evaluated at $\Theta=0\degree$ which is $O(1)$ and the remaining $\V - 1$ vertices (up to $\Theta=360\degree$) are determined using the binary search procedure. Thus, the time complexity of the algorithm is $O((\V-1)\log n) \simeq O(\V \log n)$. The search domain $n = (\theta_2 - \theta_1)$ is $360\degree$ at maximum, so the complexity is linear on the number of vertices of the projected polytope, $O(\V)$.

%% file: experiments.tex
\section{Experiments}\label{sec:experiments}
\begin{figure}[!thb]
	\centering{}
   \includegraphics[scale=0.7]{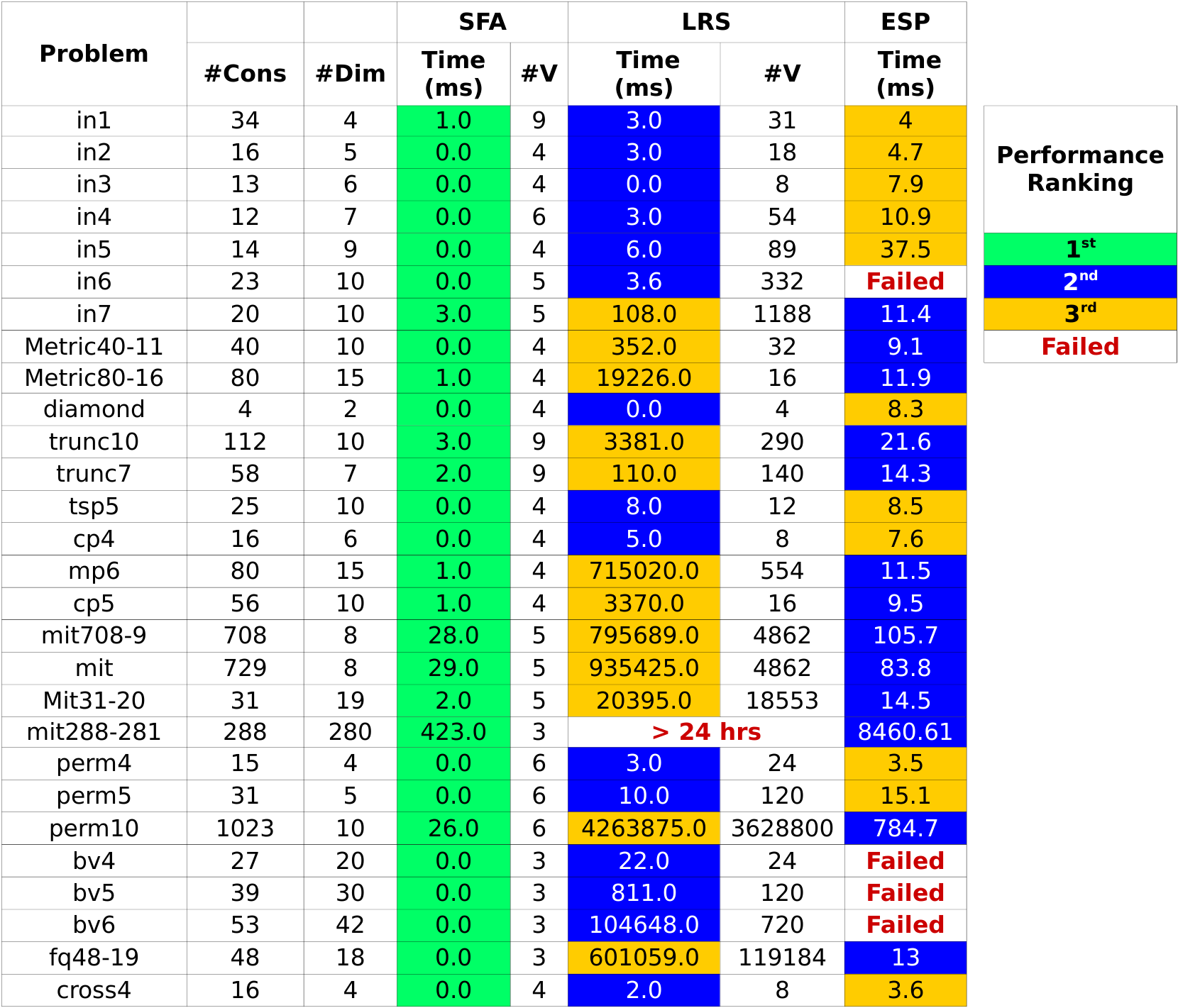}
\caption{Performance comparison of the proposed algorithm SFA with LRS and ESP. \#Cons: number of constraints, \#Dim: dimension of the polytope, \#V: number of vertices and ms: Time in millisecond.\label{fig:experiments}}
\end{figure}

We conducted the performance experiments on Intel i7-4770, 3.40GHz processor with 8GB RAM. %The performance reported is averaged over 10 runs.%
We used the benchmarks reported in (\url{http://cgm.cs.mcgill.ca/~avis/C/lrslib/archive/}) \cite{avis1998computational}. Figure \ref{fig:experiments} reports the performance of our proposed support-function-based algorithm (referred as SFA) with $\epsilon=0.5$, in comparison to vertex enumeration algorithm LRS\cite{avislrs} and projection algorithm ESP\cite{jones2004equality}. Figure \ref{fig:saf-mit-plot} and Figure \ref{fig:esp-plot} shows the 2D projected polytope of a benchmark using our algorithm and using the ESP algorithm respectively, illustrating correctness of our algorithm.
We see that the number of vertices in a 2D projection of a polytope is significantly less as compared to the vertices for the polytope. For instance, the polytope \textbf{perm10} have $3628800$ vertices, whereas a 2D projection has only 6 vertices. Our algorithm took only 26 milliseconds to enumerate all the vertices of a 2D projection whereas LRS took more that one hour for the enumeration and the ESP took 784.7 milliseconds to just perform the projection on the same polytope on the same dimensions. 
%We further observed that for very large model such as \textbf{mit288-281} both the methods lrs and pd failed to enumerate on our above mentioned experimental setup, reporting the error message \emph{Digits is too large}, where as using our approach it enumerated all the $10$ vertices in just $1316.3$ milliseconds ($1.31$ seconds). We also observed that for most models lrs performs better than pd. In addition, pd fails to enumerate if the origin is not in the interior of the polytope, which is not the case in the proposed algorithm.
Experiments shows a better performance of SFA on all benchmarks as compared to ESP and LRS. 

%% file: conclusion.tex
\section{Conclusion}\label{conclude}
We propose an efficient vertex enumeration algorithm of a two-dimensional projection of H-polytope. The proposed algorithm performs better than the existing approaches of enumerating vertices of a projection. The generated vertices are sorted in a counter-clockwise order and works for any H-polytope.